\documentclass{ws-procs9x6-cpt19}
\usepackage{tikz}

\begin{document}

\newcommand{\refeq}[1]{(\ref{#1})}
\def\etal {{\it et al.}}


\def\al{\alpha}
\def\be{\beta}
\def\ga{\gamma}
\def\de{\delta}
\def\ep{\epsilon}
\def\ve{\varepsilon}
\def\ze{\zeta}
\def\et{\eta}
\def\th{\theta}
\def\vt{\vartheta}
\def\io{\iota}
\def\ka{\kappa}
\def\la{\lambda}
\def\vpi{\varpi}
\def\rh{\rho}
\def\vr{\varrho}
\def\si{\sigma}
\def\vs{\varsigma}
\def\ta{\tau}
\def\up{\upsilon}
\def\ph{\phi}
\def\vp{\varphi}
\def\ch{\chi}
\def\ps{\psi}
\def\om{\omega}
\def\Ga{\Gamma}
\def\De{\Delta}
\def\Th{\Theta}
\def\La{\Lambda}
\def\Si{\Sigma}
\def\Up{\Upsilon}
\def\Ph{\Phi}
\def\Ps{\Psi}
\def\Om{\Omega}
\def\mn{{\mu\nu}}

\def\cL{{\cal L}}

\def\fr#1#2{{{#1} \over {#2}}}
\def\half{{\textstyle{1\over 2}}}

\def\prt{\partial}

\def\etal{{\it et al.}}

\def\pt#1{\phantom{#1}}
\def\ol#1{\overline{#1}}

\def\sb{\overline{s}{}}

\def\stx{\sb^{TX}}
\def\sty{\sb^{TY}}
\def\stz{\sb^{TZ}}

\newcommand{\beq}{\begin{equation}}
\newcommand{\eeq}{\end{equation}}
\newcommand{\bea}{\begin{eqnarray}}
\newcommand{\eea}{\end{eqnarray}}
\newcommand{\bit}{\begin{itemize}}
	\newcommand{\eit}{\end{itemize}}
\newcommand{\rf}[1]{(\ref{#1})}

\title{Ring Laser Gyroscope Tests of Lorentz Symmetry}

\author{Max L. Trostel,$^1$ Serena Moseley,$^1$ Nicholas Scaramuzza,$^2$ and Jay D.\ Tasson$^1$}

\address{$^1$Physics and Astronomy Department, Carleton College, Northfield, MN 55057, USA\\
$^2$Physics Department, St.\ Olaf College, Northfield, MN 55057, USA}

\begin{abstract}
Interferometric gyroscope systems are being developed 
with the goal of measuring general-relativistic effects 
including frame-dragging effects. 
Such devices are also capable of performing searches for Lorentz violation. 
We summarize efforts that relate gyroscope measurements 
to coefficients for Lorentz violation 
in the gravity sector of the Standard-Model Extension.
\end{abstract}

\bodymatter

\section{Interferometric gyroscopes and Lorentz violation}
Lorentz violation in the Standard-Model Extension (SME)
can be sought\cite{moseley} 
using interferometric gyroscopes 
based on light\cite{light} or matter\cite{matter} waves.  
Here, we summarize some results from Ref.\ \refcite{moseley}
with a focus on light-based systems, 
which consist of beams traveling around a closed path in opposite directions.  
Effects that break the symmetry of the counter-propagating beams 
are encoded in their interference.
The largest such effect routinely observed,
the Sagnac effect, 
is due to rotation, 
which generates a beat frequency
$
\nu_{\rm s} = \frac{4 A \vec{\Omega} \cdot{\hat{n}}}{P \lambda},
$
where $A$ is the area enclosed, 
$\vec{\Omega}$ is the angular velocity, 
$\hat{n}$ is the vector normal to the loop, 
$P$ is its perimeter, 
and $\lambda$ is the wavelength of the light.
A fixed system on Earth will experience several effects 
that alter the beat frequency, 
including the Sagnac effect of Earth's rotation 
at angular frequency $\om$ 
and the general-relativistic frame-dragging effect. 
Lorentz violation as described by the SME\cite{sme,sme2} 
can also break the symmetry.

The contributions
to a post-Newtonian expansion of the metric $g_{\mu \nu}$ in the SME\cite{qbak} 
that are relevant for our analysis take the form
\bea
g_{0j} &=& -\sb^{0j}U - \sb^{0k} U^{jk} + \half {\hat Q}^j \ch,
\label{sme}
\eea
where $U$ is the Newtonian potential, 
$\ch$ is the superpotential,\cite{qbak} 
$U^{jk}$ is an additional post-Newtonian potential,\cite{qbak} 
and $\sb^\mn$ is the $d=4$ coefficient for Lorentz violation 
that provides the relevant minimal effects.  
The $d=5$ effects are contained in 
${\hat Q}^j = [q^{(5) 0jk0l0m}+q^{(5) n0knljm} +q^{(5) njknl0m}] \prt_k\prt_l\prt_m$ 
where $q^{(5) \mu \rh \al \nu \be \si \ga}$ 
is the coefficient for Lorentz violation.\cite{kmgw}

\section{Measuring Lorentz violation}
\label{sec:sig}
\begin{figure}\centering
\begin{tikzpicture}
\clip (-4,3.5) rectangle + (8,-4.6);
	\shade[ball color = gray!40, opacity = 0.4] (0,0) circle (2.5cm);
	\draw (0,0) circle (2.5cm);
	\draw (-2.5,0) arc (180:360:2.5 and 1);
	\draw[dashed] (2.5,0) arc (0:180:2.5 and 1);

	\draw[thick] (-0.45,-0.45) arc (250:332:1.25 and 0.5) node [xshift=-0.7cm,yshift=-0.45cm] {$\ph$};
	\draw[thick] (0.65,0.65) arc (43:79:1.2 and 0.5) node [xshift=0.4cm,yshift=0.10cm] {$\th$};

	\fill[fill=black] (0,0) circle (1.5pt);
	\fill[fill=black] (0,2.5) circle (1.5pt);
	\fill[fill=black] (1.695,1.695,0.707) circle (1.5pt) ;
	\draw[dashed] (0, 0, 0) -- (1.695,0,0.707);
	\draw[dashed] (1.695,0,0.707) -- (1.695,1.695,0.707);
	
	\draw[thick] (0,0) -- (1.695,1.695,0.707);

	\draw[-latex, thick] (0,2.5,0) -- (0,3.0,0) node[left]{\large $Z$ };
	\draw[-stealth, thick] (-0.2,2.8,0) arc (-240:60:0.4 and 0.1) node[right]{\large $\phantom{\om}\om$ };
	
	\draw [->] (0,0,0) -- (0,0,1.6) node [xshift=-0.19cm,yshift=-0.09cm] {$X$};
	\draw [->] (0,0,0) -- (0,1.6,0) node [xshift=0.19cm,yshift=-0.08cm] {$Z$};
	\draw [->] (0,0,0) -- (1.6,0,0) node [xshift=0.16cm] {$Y$};

	\draw [->] (1.695,1.695,0.707) -- (2.311,1.695,-0.767) node [xshift=0.55cm] {$y$ (East)};
	\draw [->] (1.695,1.695,0.707) -- (2.78,2.78,1.16) node [xshift=0.45cm,yshift=0.11cm] {$z$ (Up)};
	\draw [->] (1.695,1.695,0.707) -- (2.697,0.519,1.125) node [xshift=0.62cm,yshift=-0.1cm] {$x$ (South)};
	\draw [thick, -latex] (1.695,1.695,0.707) -- (3.645, 1.629, 0.269) node [xshift=0.15cm,yshift=0.09cm] {$\hat n$};
	\draw [dashed] (1.695,1.695,0.707) -- (3.12465, 0.655756, -0.228985);
	\draw [dashed] (3.12465, 0.655756, -0.228985) -- (3.645, 1.629, 0.269);
	
	\draw[thick] (2.07,0.4) arc (313:349:2 and 1) node [xshift=-0.05cm,yshift=-0.39cm] {$\beta$};
	\draw[thick] (2.51,1.47) arc (5:36:2.9 and 1) node [xshift=0.62cm,yshift=-0.25cm] {$\alpha$};

	\end{tikzpicture}
	\caption{Location of the laboratory
in shifted Sun-centered frame\cite{data} axes $(X, Y, Z)$
and orientation of the normal vector of the gyroscope $\hat{n}$
relative to laboratory coordinates $(x,y,z)$.}
\label{fig:earth}
\end{figure}
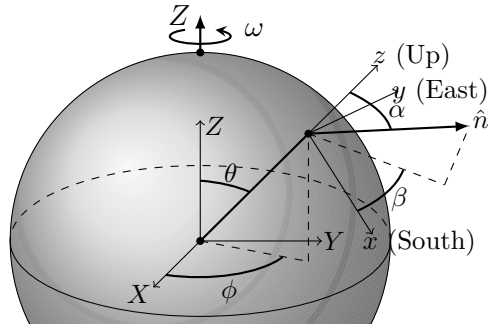

The beat frequency measured by a light-based gyroscope\cite{bosi} 
is related to the proper time difference in the lab $\Delta \tau$ 
between the paths taken by the counter-propagating beams 
via $\nu = \frac{\De \tau}{\la P}$,
and at leading order in Lorentz violation 
we find in our post-Newtonian expansion:
	\beq
	\De \tau \approx 2 \oint{ g_{0j} dx^j},
	\label{det}
	\eeq
where the integral is taken around the closed interferometer loop.
By analogy with Amp\`ere's law, 
we transform this line integral into an integration over area, 
which simplifies into a product. 
In order to derive a general result 
that is valid for any Earth-based laboratory 
and for any ring orientation within that laboratory, 
we use the angles 
$\theta$, $\phi$, $\alpha$, and $\beta$ 
shown in Fig.\ \ref{fig:earth}.
Given $d = 4$ coefficients for Lorentz violation, 
we find a beat frequency\cite{moseley}
\bea
\nonumber
\nu^{(4)}_{\rm LV} &=&
\frac{4 A G M}{\la P R^2}  \sin \al 
[ \cos \be ( \stx \sin \ph - \sty \cos \ph) \\
&&
+ \sin \be ( \cos \th ( \stx \cos \ph + \sty \sin \ph) - \stz \sin \th) ],
\label{fblv}
\eea
where $M$ and $R$ are Earth's mass and radius, 
respectively.
The small terms suppressed by an Earth-revolution boost factor have been omitted.
A sample special case of this result is found in Ref.\ \refcite{ns}.
For $d=5$ coefficients, 
we find a beat frequency with a similar form
(omitted here for brevity),
which can be written in terms of 
the 15 canonical coefficient combinations 
written as components of $K_{JKLM}$.\cite{pulsar} 
The angle $\phi$ varies at the sidereal frequency, 
which indicates that both dimension four and five coefficients for Lorentz violation 
will produce signals with this time dependence.

\section{Experiments}
A number of planned or ongoing experiments 
may be of interest in the context of this work.
For example, 
the Gyroscopes IN GEneral Relativity (GINGER) experiment, 
which is designed to measure 
the de Sitter and Lense--Thirring effects of General Relativity,\cite{lasgyro,bosi} 
expects to obtain sensitivities to the angular velocity of the Earth 
via the Sagnac effect beyond the part in $10^{9}$ level 
reaching perhaps a few parts in $10^{12}$.\cite{lasgyro} 
It is possible to generate crude estimates of the sensitivity to Lorentz violation 
that might be expected using these goals 
and the results outlined in Sec.\ \ref{sec:sig}.  
We find $\sb^{TJ} \approx \frac{\ep \om c R^2}{G M} \approx 10^{-6}$ 
and $K_{JKLM} \approx \frac{\ep \om c R^3 }{G M} \approx 10\,$m
in SI units with $\ep=10^{-9}$ as the fractional sensitivity to $\om$ 
and $c$ as the speed of light.  
For $\bar{s}^{TJ}$, 
these sensitivities are competitive with other laboratory experiments,\cite{data} while for dimension five coefficients, 
these sensitivities are competitive with the best existing measurements 
from binary pulsars.\cite{pulsar}

\section*{Acknowledgments}
The authors gratefully acknowledge financial support as follows:
S.M.\ from Carleton Summer Science Fellows,
N.S.\ from the St.\ Olaf CURI fund.


\begin{thebibliography}{xx}
\bibitem{moseley}
S.\ Moseley, N. Scaramuzza, J.D.\ Tasson, and M.L.\ Trostel, 
arXiv:1907.05933.

\bibitem{light}
See, for example,
N.\ Beverini \etal, 
J.\ Phys.\ Conf.\ Ser.\ {\bf 723}, 012061 (2016);
A.D.V.\ Di Virgilio,
these proceedings, arXiv:1906.04156.

\bibitem{matter}
See, for example,
D.\ Savoie \etal, 
Sci.\ Adv.\ {\bf 4}, eaau7948 (2018).

\bibitem{sme}
D.\ Colladay and V.A.\ Kosteleck\'y,
Phys.\ Rev.\ D {\bf 58}, 116002 (1998);
V.A.\ Kosteleck\'y,
Phys.\ Rev.\ D {\bf 69}, 105009 (2004).

\bibitem{sme2}
For a review, 
see J.D.\ Tasson, 
Rept.\ Prog.\ Phys.\ {\bf 77}, 062901 (2014).

\bibitem{qbak}
Q.G.\ Bailey and V.A.\ Kosteleck\'y, 
Phys.\ Rev.\ D {\bf 74}, 045001 (2006).
	
\bibitem{kmgw}
Q.G.\ Bailey and D.\ Havert, 
Phys.\ Rev.\ D {\bf 96}, 064035 (2017).

\bibitem{bosi}
F.\ Bosi \etal, 
Phys.\ Rev.\ D {\bf 84}, 122002 (2011).

\bibitem{ns}
N.\ Scaramuzza and J.D.\ Tasson, 
in V.A.\ Kosteleck\'y, ed.,
{\it CPT and Lorentz Symmetry VII}, 
World Scientific, Singapore 2017.

\bibitem{pulsar}
L.\ Shao and Q.G.\ Bailey, 
Phys.\ Rev.\ D {\bf 98}, 084049 (2018).

\bibitem{lasgyro}
A.\ Ortolan \etal, 
J.\ Phys.\ Conf.\ Ser.\  {\bf 718}, 072003 (2016);
A.D.V.\ Di Virgilio \etal,
Eur.\ Phys.\ J.\ Plus {\bf 132}, 157 (2017).

\bibitem{data}
{\it Data Tables for Lorentz and CPT Violation,}
V.A.\ Kosteleck\'y and N.\ Russell,
2019 edition,
arXiv:0801.0287v12.

\end{thebibliography}
\end{document}